\documentstyle[12pt,twoside,fleqn,espcrc1]{article}
\input epsf

\newcommand{\AmS}{{\protect\the\textfont2
  A\kern-.1667em\lower.5ex\hbox{M}\kern-.125emS}}

\title{$J/\psi$ and $\psi^\prime$ suppression by comovers in Pb+Pb collisions}

\author{Sean Gavin\address{Physics Department, Brookhaven National
Laboratory, Upton, NY}
and 
Ramona Vogt\address{Physics Department, Lawrence Berkeley National 
Laboratory, Berkeley, CA}%
\address{Physics Department, University of California, Davis, CA}%
\thanks{This manuscript has been authored under contracts
DE-AC02-76CH00016 and DE-AC03-76SF0098 with the U. S. Department of Energy.}}

\begin{document}
\maketitle

\begin{abstract}
Measurements of $\psi$ and $\psi^\prime$ production at the CERN SPS
are compared to predictions based on a hadronic model of charmonium
suppression.  Detailed information is presented to facilitate
comparison to other analyses.  Sensitivity of these conclusions to the
model parameters is discussed.
\end{abstract}

\section{INTRODUCTION}
Has the quark--gluon plasma been discovered at the SPS?  The NA50
collaboration reported a dramatic decrease in $\psi$ production in
Pb+Pb collisions at 158 GeV per nucleon \cite{na50}.  Specifically, M.
Gonin presented at this meeting a striking `threshold effect' in the
$\psi$--to--continuum ratio as a function of a calculated quantity,
the mean path length of the $\psi$ through the nuclear medium, $L$,
see fig.~1.  This apparent threshold has sparked considerable
excitement as it may signal deconfinement in the heavy Pb+Pb system
\cite{bo}.

In this talk we report on work in ref.~\cite{gv2} comparing Pb results
to predictions using a hadronic model of charmonium suppression.  We
first demonstrate that the behavior in the NA50 plot is not a
threshold effect but, rather, reflects the approach to the geometrical
limit of $L$ as the symmetric Pb+Pb collisions become increasingly
central.  When plotted as a function of the {\it measured} neutral
transverse energy $E_{T}$ as in fig.~2, the data varies smoothly as in
S+U measurements in fig.~4b below \cite{na38,na38c,na38d,na38e}.  The
difference between S+U and Pb+Pb data lies strictly in the relative
magnitude.  To assess this magnitude, we compare $\psi$ and
$\psi^\prime$ data to expectations based on the hadronic comover model
\cite{gv,gstv}.  The curve in fig.~1 represents our calculation using
parameters fixed earlier in Ref.\ \cite{gstv}.  Our result is
essentially the same as the Pb+Pb prediction in \cite{gv}.
\begin{figure}[htb]
\epsfysize=5in
\centerline{\epsffile{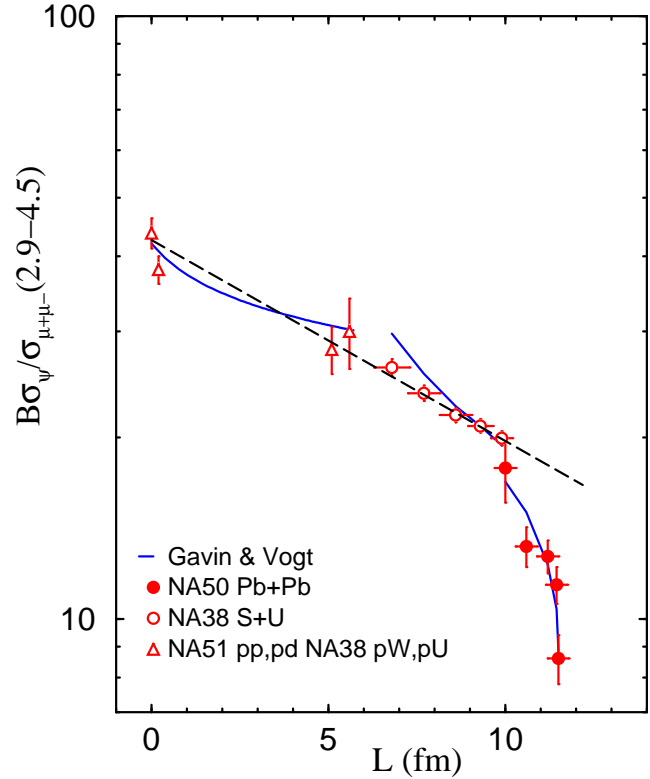}}
\caption[]{NA50 comparison [1] of Pb+Pb and S+U$\rightarrow \psi +X$ as
functions of the average path length $L$, see eq.\ (3).  $B$ is the
$\psi\rightarrow \mu^+\mu^-$ branching ratio.}
\label{fig:fig1}
\end{figure}
\begin{figure}[htb]
\epsfysize=5in
\centerline{\epsffile{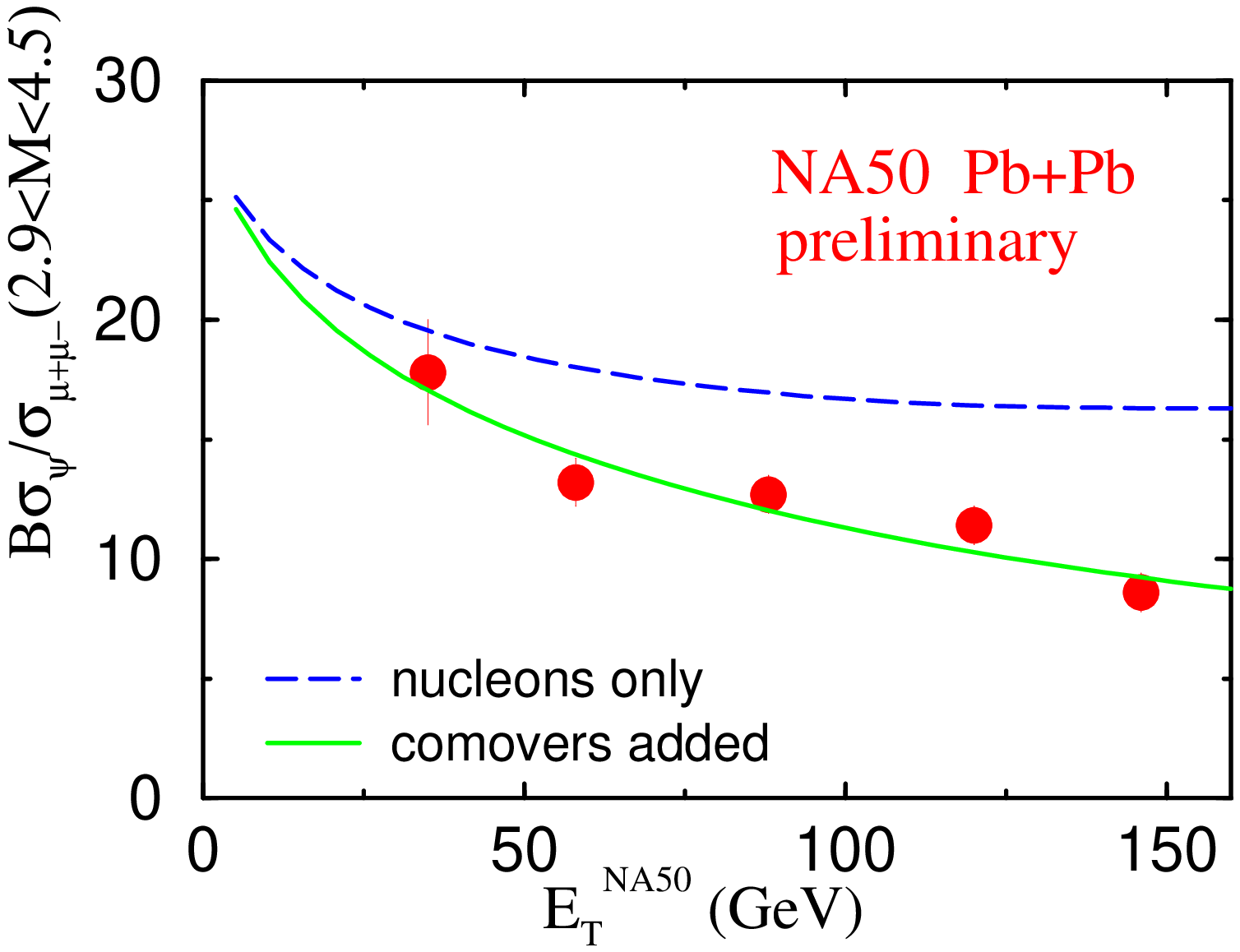}}
\caption{$\psi$--to--continuum ratio as a function of $E_T$ from [1] 
compared to the comover prediction [3].}
\label{fig:fig2}
\end{figure}

Our primary intention in this work is to demonstrate that there is no
evidence for a strong discontinuity between $p$A, S+U and Pb+Pb data.
Secondly, we show that our model predictions agree with the new Pb+Pb
data.  The consistency of our predictions can be judged by comparing
our old $p$A and S+U calculations with more recent NA38 and NA51 data.
However, we must be cautious about taking model details too seriously.
Regarding the first point, we quote Maurice Goldhaber: ``$\ldots$
absence of evidence is {\it not} evidence of absence.''  Moreover, we
do not attempt to show that our comover interpretation of the data is
unambiguous -- this is certainly impossible at present.  Finally, the
significance of our present results must be weighted by the fact that
all $p$A and AB data are preliminary and at different beam energies.
Whether the newer data is indeed `improved' has yet to be established.
One should be particularly cautious in taking literally the values of
model parameters, such as the comover density.

\section{NUCLEONS AND COMOVERS}
The hadronic contribution to the suppression arises from scattering of
the nascent $\psi$ with produced particles -- the comovers -- and
nucleons \cite{gv2,gv,gstv}.  To determine the suppression from nucleon
absorption, we calculate the probability that a $c{\overline c}$ pair
produced at a point $(b, z)$ in a nucleus survives scattering with
nucleons to form a $\psi$.  The standard \cite{gstv,gh} result is
\begin{equation}
S_{A} = {\rm exp}\{-\int_z^\infty\! dz\, \rho_{A}(b, z) \sigma_{\psi N}\}
\end{equation}
where $\rho_{A}$ is the nuclear density, $b$ the impact parameter and
$\sigma_{\psi N}$ the absorption cross section for $\psi$--nucleon
interactions.  One can estimate $S_{A}\sim \exp\{-
\sigma_{\psi N} \rho_0 L_{A}\}$, where $L_{A}$ is the path length
traversed by the $c\overline{c}$ pair.  On average, $L_{A}$ cannot
exceed the nuclear radius $R_{A}$, since the pair can be produced
anywhere within the nucleus.

Suppression can also be caused by scattering with mesons that happen
to travel along with the $c\overline{c}$ pair (see refs.\ in
\cite{gv}).  The density of such comovers scales roughly with $E_{T}$.
The corresponding survival probability is
\begin{equation}
S_{\rm co} = {\rm exp}\{- \int\! d\tau n\, 
\sigma_{\rm co} v_{\rm rel}\},
\end{equation}
where $n$ is the comover density and $\tau$ is the time in the $\psi$
rest frame.  We write $S_{\rm co}\sim {\rm exp}\{-\beta E_{T}\}$,
where $\beta$ depends on the scattering frequency,
the formation time of the comovers and the transverse size of the
central region, $R_{T}$, {\it cf.} eq.\ (8).

To understand the saturation of the Pb data with $L$ in fig.~1, we apply
the schematic approximation of Ref.~\cite{gh} for the moment to write
\begin{equation}
{{\sigma^{AB}_\psi(E_{T})}\over{\sigma^{AB}_{\mu^+\mu^-}(E_{T})}}
\propto \langle S_{A}S_{B}S_{\rm co}\rangle
\sim 
{\rm e}^{-\sigma_{\psi N}\rho_{0}L}{\rm e}^{-\beta E_{T}},
\end{equation}
where the brackets imply an average over the collision geometry for
fixed $E_{T}$ and $\sigma(E_T) \equiv d\sigma/dE_T$.  The path length
$L\equiv \langle L_{A}+L_{B}\rangle$ and transverse size $R_T$ depend
on the collision geometry.  The path length grows with $E_{T}$, but
most stop at the geometric limit $R_A + R_B$ in central collisions.
Explicit calculations show that nucleon absorption begins to {\it
saturate} for $b < R_A$, where $R_A$ is the smaller of the two nuclei, 
see fig.~6 below.  On the other hand, particle production and,
consequently, $E_{T}$ continue to grow after $L$ saturates due, {\it
e.g.}, to fluctuations in the number and hardness of $NN$ collisions.
Equation (2) falls exponentially in the `super--central' regime,
$b < R_A$, because $\beta$ is essentially fixed. 

Important for understanding the NA50 results is the fact that
saturation occurs at larger values of the average impact parameter for
Pb+Pb system compared to S+U, see fig.~6 below.  This result reflects
the striking geometric difference between symmetric and asymmetric
nuclear collisions.  Specifically, in the S+U system there is large
region, $R_{\rm S} < b < R_{S} + R_{\rm U}$, where varying $b$ changes
the volume of the interaction region dramatically.  In contrast, $R_A
= R_B$ in Pb+Pb collisions and $R_{\rm Pb}\approx 6.6$~fm is large.  We
show below that typical values of $b$ corresponding to the $E_T$
region covered by NA50 data do not go far into the region $R_{\rm Pb}
< b < 2R_{\rm Pb}$ where [The importance of $b$ being near zero for
saturation to occur was overemphasised in SG's oral presentation of
this talk.]

\section{$J/\psi$ SUPPRESSION}
In fig.~1, we compare the Pb data to calculations of the
$\psi$--to--continuum ratio that incorporate nucleon and comover
scattering.  The contribution due to nucleon absorption indeed levels
off for small values of $b$, as expected from eq.\ (3).  Comover
scattering accounts for the remaining suppression.

These results are {\it predictions} obtained using the computer code
of Ref.~\cite{gv} with parameters determined in Ref.~\cite{gstv}.
However, to confront the present NA50 analysis \cite{na50}, we must
perform the following updates:
\begin{itemize}
\item Calculate the continuum dimuon yield in the new mass range $2.9
< M < 4.5$~GeV.  
\item Adjust the $E_T$ scale to the pseudorapidity
acceptance of the NA50 calorimeter, $1.1 < \eta < 2.3$.
\end{itemize} 
The agreement in fig.~1 depends strongly on these adjustments.

We now review the details of our calculations, highlighting the
adjustments as we go.  For collisions at a fixed $b$, the
$\psi$--production cross section is
\begin{equation}
\sigma_\psi^{AB}(b)
=
\sigma^{NN}_{\psi}\!\int\! d^2s dz dz^\prime\,\rho_A(s,z)
\rho_B(b-s,z^\prime)\, S 
\end{equation}
where $S\equiv S_AS_BS_{\rm co}$ is the product of the survival
probabilities in the projectile $A$, target $B$ and comover matter.
The continuum cross section is
\begin{equation}
\sigma_{\mu^{+}\mu^{-}}^{AB}(b) = 
\sigma^{NN}_{\mu^+\mu^-}\!\int\! d^2s dz dz^\prime\,\rho_A(s,z)
\rho_B(b-s,z^\prime).
\end{equation}
The magnitude of eqs.\ (4), (5) and their ratio is fixed by the
elementary cross sections $\sigma^{NN}_{\psi}$ and
$\sigma^{NN}_{\mu^{+}\mu^{-}}$.  We calculate $\sigma^{NN}_{\psi}$
using the phenomenologically--successful color evaporation model
\cite{hpc-psi}.  The continuum in the mass range used by NA50, $2.9 <
M < 4.5$~GeV, is described by the Drell--Yan process \cite{hpc-dy}.
To confront NA50 and NA38 data in the appropriate kinematic
regime, we compute these cross sections at leading order using GRV LO
parton distributions with a charm $K$--factor $K_c= 2.7$ and a color
evaporation coefficient $F_\psi =2.54\%$ from \cite{hpc-psi} and a
Drell--Yan $K$--factor $K_{DY}=2.4$.  Using these cross sections to
construct the normalized ratios in fig.~2 corresponds to the first
update.

To obtain $E_T$ dependent cross sections from eqs.\ (4) and (5), we
write
\begin{equation}
\sigma^{AB}(E_{T}) =
\int\! d^2b\, P(E_T,b) \sigma^{AB}(b).
\end{equation}
The probability $P(E_T,b)$ that a collision at impact parameter $b$
produces transverse energy $E_T$ is related to the minimum--bias
distribution by
\begin{equation}
\sigma_{\rm min}(E_{T}) = \int\! d^{2}b\; P(E_{T}, b).
\end{equation}
We parametrize $P(E_{T}, b) = C\exp\{- (E_{T}- {\overline
E}_{T})^2/2\Delta\}$, where ${\overline E}_{T}(b) = \epsilon {\cal
N}(b)$, $\Delta(b) = \omega \epsilon {\overline E}_{T}(b)$,
$C(b)=(2\pi\Delta(b))^{-1}$ and ${\cal N}(b)$ is the number of
participants (see, {\it e.g.}, Ref.~\cite{gv}).
We take $\epsilon$ and $\omega$ to be phenomenological
calorimeter--dependent constants.

We compare the minimum bias distributions for total hadronic $E_T$
calculated using eq.\ (7) for $\epsilon = 1.3$~GeV and $\omega = 2.0$
to NA35 S+S and NA49 Pb+Pb data \cite{na49}.  The agreement in fig.~3a
builds our confidence that eq.\ (7) applies to the heavy Pb+Pb system.
\begin{figure}[htb]
\vskip -1.5truein
\epsfysize=5.5in
\centerline{\epsffile{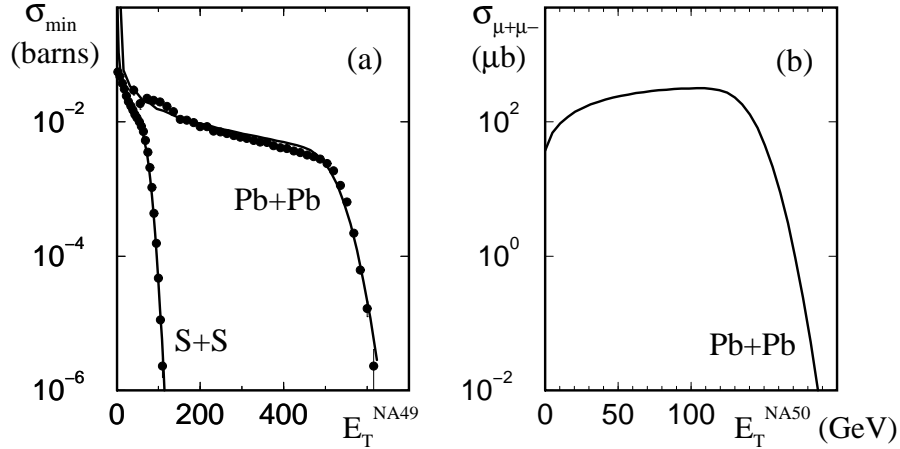}}
\vskip -1.0truein
\caption{Transverse energy distributions from eqs.\ (7, 8).
The S--Pb comparison (a) employs the same parameters.}
\end{figure}
Figure 3b shows the distribution of neutral transverse energy
calculated using eqs.\ (5) and (6) to simulate the NA50 dimuon
trigger. We take $\epsilon = 0.35$~GeV, $\omega = 3.2$, and
$\sigma^{NN}_{\mu^+\mu^-}\approx 35.2$~pb as appropriate for the
dimuon--mass range $2.9 < M < 4.5$~GeV.  The $E_T$ distribution for
S+U~$\rightarrow \mu^+\mu^- + X$ from NA38 was described \cite{gstv}
using $\epsilon = 0.64$~GeV and $\omega = 3.2$ -- the change in
$\epsilon$ corresponds roughly to the shift in particle production
when the pseudorapidity coverage is changed from $1.7 < \eta < 4.1$
(NA38) to $1.1 < \eta < 2.3$ (NA50).  Taking $\epsilon = 0.35$~GeV is
the second update listed earlier.  Though not shown here, we comment
that the calculations agree with the NA50 data at high $E_{T}$, but
differ somewhat at low $E_{T}$ due to the efficiency of the active
target, which is lower for more peripheral collisions.

We now apply eqs.\ (1,2,4) and (5) to charmonium suppression in Pb+Pb
collisions.  To determine nucleon absorption, we used $p$A data to fix
$\sigma_{\psi N}\approx 4.8$~mb in Ref.~\cite{gstv}. This choice is in
accord with the latest NA38 and NA51 $pA$ data, see fig.~4a.
\begin{figure}[htb]
\vskip -2.0truein
\epsfysize=5.0in
\centerline{\epsffile{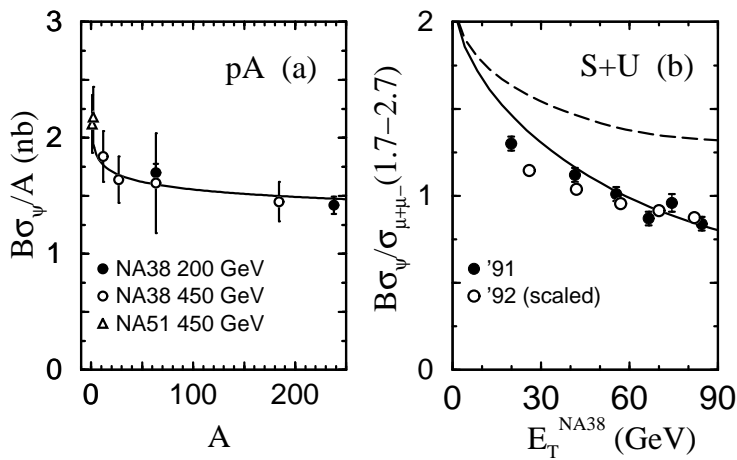}}
\vskip -0.5truein
\caption{(a) $p$A cross sections [1] in the
NA50 acceptance and (b) NA38 S+U ratios from the '91 [6] and '92 [1]
runs. '92 data are scaled to the '91 continuum.  The dashed line
indicates the suppression from nucleon absorption alone.}
\end{figure}
To specify comover scattering \cite{gstv}, we assumed that the
dominant contribution to $\psi$ dissociation comes from exothermic
hadronic reactions such as $\rho + \psi \rightarrow D+ \overline{D}$.
We further took the comovers to evolve from a formation time
$\tau_{0}\sim 2$~fm to a freezeout time $\tau_{F}\sim R_{T}/v_{\rm
rel}$ following Bjorken scaling, where $v_{\rm rel}\sim 0.6$ is
roughly the average $\psi-\rho$ relative velocity.  The
survival probability, eq.\ (2), is then
\begin{equation}
S_{\rm co} = \exp\{ - \sigma_{\rm co}v_{\rm rel}n_{0}\tau_{0} 
\ln(R_{T}/v_{\rm rel}\tau_{0})\}
\end{equation}
where $\sigma_{\rm co} \approx 2\sigma_{\psi N}/3$, $R_{T}\approx
R_{A}$ and $n_{0}$ is the initial density of sufficiently massive
$\rho, \omega$ and $\eta$ mesons.  To account for the variation of
density with $E_{T}$, we take $n_{0} = {\overline
n}_{0}E_{T}/{\overline E}_{T}(0)$ \cite{gv}. A value $\overline{n}_{0}
= 0.8$~fm$^{-3}$ was chosen to fit the central S+U datum.  Since we
fix the density in central collisions, this simple {\it ansatz} for
$S_{\rm co}$ may be inaccurate for peripheral collisions.

We expect the comover contribution to the suppression to increase in
Pb+Pb relative to S+U for central collisions because both the
initial density and lifetime of the system can increase.  To be
conservative, we assumed that Pb and S beams achieve the same mean initial
density.  Even so, the lifetime of the system essentially doubles in
Pb+Pb because $R_T \sim R_{A}$ increases to 6.6~fm from 3.6~fm in S+U.
The increase in the comover contribution evident in comparing figs.~2
and 4b is described by the seemingly innocuous logarithm in eq.\ (8),
which increases by $\approx 60\%$  in the larger Pb system.

In Ref.~\cite{gstv}, we pointed out that comovers were necessary to
explain S+U data from the NA38 1991 run \cite{na38}.  Data just
released \cite{na50} from their 1992 run support this conclusion.  The
'91 $\psi$ data were presented as a ratio to the dimuon continuum in
the low mass range $1.7 < M < 2.7$~GeV, where charm decays are an
important source of dileptons.  On the other hand, the '92 $\psi$ data
\cite{na50,na38e} are ratios to the Drell--Yan cross section in the
range $1.5< M < 5.0$~GeV.  That cross section is extracted from the
continuum by fixing the $K$--factor in the high mass
region \cite{na38f}.  To compare our result from Ref.~\cite{gstv} to
these data, we scale the '92 data by an empirical factor.  This factor
is $\approx 10\%$ larger than our calculated factor
$\sigma^{NN}_{DY}(92)/\sigma^{NN}_{\rm cont.}(91) \approx 0.4$; these
values agree within the NA38 systematic errors.  [NA50 similarly
scaled the '92 data to the high--mass continuum to produce fig.~1.]
Because our fit is driven by the highest $E_T$ datum, we see from
fig.~4b that a fit to the '92 data would not appreciably change our
result.  Note that a uniform decrease of the ratio by 10\% would
increase the comover contribution needed to explain S+U collisions.

NA50 and NA38 have also measured the total $\psi$--production cross
section in Pb+Pb \cite{na50} and S+U reactions \cite{na38c}.  To
compare to that data, we integrate eqs.\ (4, 6) to obtain the total
$(\sigma/AB)_{\psi} = 0.97$~nb in S+U at 200~GeV and 0.54~nb for Pb+Pb
at 158~GeV in the NA50 spectrometer acceptance, $0.4 > x_{F}> 0$ and
$-0.5 < \cos\theta < 0.5$ (to correct to the full angular range and $1
> x_{F} > 0$, multiply these cross sections by $\approx 2.07$).  The
experimental results in this range are $1.03 \pm 0.04 \pm 0.10$~nb for
S+U collisions \cite{na38} and $0.44 \pm 0.005 \pm 0.032$ nb for Pb+Pb
reactions \cite{na50}.  Interestingly, in the Pb system we find a
Drell--Yan cross section $(\sigma/AB)_{{}_{DY}} = 35.2$~pb while NA50
finds $(\sigma/AB)_{{}_{DY}} = 32.8\pm 0.9\pm 2.3$~pb.  Both the
$\psi$ and Drell--Yan cross sections in Pb+Pb collisions are somewhat
above the data, suggesting that the calculated rates at the $NN$ level
may be $\sim 20-30\%$ too large at 158~GeV.  Such an error is
consistent with ambiguities in current $pp$ data near that low energy
\cite{hpc-psi}.  Note also that nuclear effects on the parton
densities omitted in eqs.\ (4,5) can affect the total S and Pb cross
sections at this level.

\section{COMPARISON TO OTHER WORK}
\begin{figure}[htb]
\vskip -2.0truein
\epsfysize=5.0in
\centerline{\epsffile{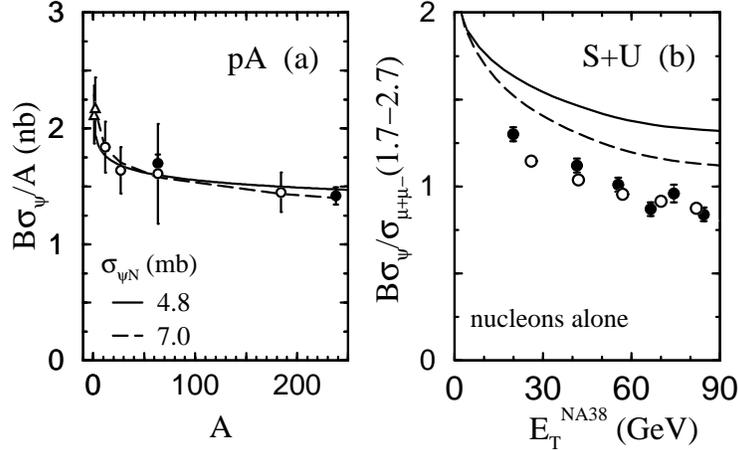}}
\vskip -0.5truein
\caption{Suppression due to nucleon absorption alone using a value
$\sigma_{\psi N} = 7$~mb as suggested by Kharzeev and Satz [2] does
not describe S+U data (b).  Both this value and our preferred 
4.8~mb are consistent with pA data (a).}
\end{figure}
To compare more easily our results with those in Refs.~\cite{na50,bo},
we present the following technical points.  Blaizot and Ollitrault
\cite{bo} and Kharzeev in these proceedings have observed that one can
describe the measured total cross sections in $p$A and S+U in the
absence of comovers if one takes $\sigma_{\psi N} = 6.2$~mb.  We also
find this to be true -- neglecting comovers and taking their value for
the absorption cross section, we obtain $(\sigma/AB)_{\psi} = 1.05$~nb
in S+U at 200~GeV and 0.62~nb for Pb+Pb at 158~GeV.  On the other
hand, our calculated $E_T$--dependent ratios do {\it not} agree with
measurements, as shown in fig.~5 (see also \cite{gstv}).  The
importance of comovers in high $E_T$ events follows from the
correlation between $E_T$, centrality and particle (comover)
multiplicity.  On the other hand, comovers only contribute to the
total S+U cross section at the $\sim 18\%$ level, because the
impact--parameter integrated cross section is dominated by large $b$
and the distinction between central and peripheral interactions is
more striking for the asymmetric S+U system.

How well does our value $\sigma_{\psi N}=$4.8~mb agree with the $p$A
data?  We estimate the $\chi^2$ per degree of freedom to be 0.34 for
$\sigma_{\psi N}=0.62$~mb and 1.3 for $\sigma_{\psi N}=0.48$~mb. Both
values of $\chi^2/{\rm d.o.f.}$ are close to unity and indicate
reasonable fits.  Given the paucity of the data and the crudeness of
the models, we consider the agreement to data to be equivalent.
Observe that including the measured S+U total cross section in
extracting $\sigma_{\psi N}$ from data as in \cite{na50} introduces an
inappropriate bias.

\begin{figure}[htb]
\vskip -1.5truein
\epsfysize=5.0in
\centerline{\epsffile{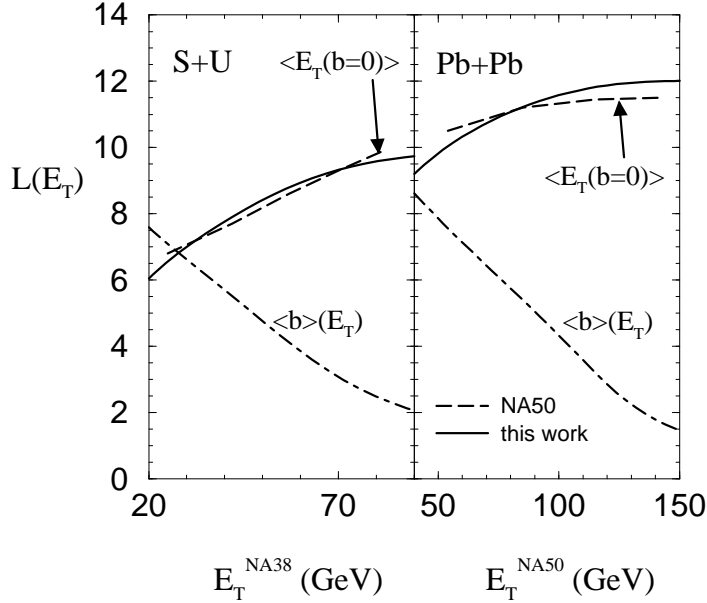}}
\vskip -0.5truein
\caption{$L(E_T)$ for S+U and Pb+Pb from (9,10) (solid curve) is in good
agreement with the NA50 result (dashed curve) used to create fig.~1
(dashed curve).  The dot--dashed curve shows the average impact
parameter $b$ obtained from $P(E_{T},b)$.  The dashed line begins
and ends at centroids of the lowest and highest $E_T$ bins measured.}
\end{figure}
Where does the $L(E_{T})$ in NA50's plot come from and how does it
approach the geometric?  To address these questions we estimate
$L(E_T)$.  Following \cite{gh} we neglect comovers and equate (3) to
our result for the $\psi$--to-- continuum ratio from (1) and (4-6).
We then expand in powers of $\sigma_{\psi N}$ and equate the terms
proportional to $\sigma_{\psi N}$ to find
\begin {equation}
\rho_0 L(E_{T}) = \{2 {\overline T}_{AB}\}^{-1} \int\! d^{2}b\;
P(E_{T}, b) \int\! d^{2}s\; \{[T_{A}(s) ]^{2}T_{B}(b-s) +
T_{A}(s)[T_{B}(b-s)]^{2}\}
\end{equation}
where
\begin{equation}
{\overline T}_{AB} \equiv \int\! d^{2}b\; P(E_{T},b) \int\!
d^{2}s\; T_{A}(s)T_{B}(b-s)
\end{equation}
and $T_{A} \equiv \int \rho_{A}dz$.  This quantity is strictly
meaningful for values of the mean free path $\rho_{0} \sigma_{\psi N}
\ll L$, as are the results of \cite{gh}.  Following \cite{na50,gh}, we
overlook the fact that this is not true for realistic parameter
values, and obtain the result in fig.~6.  We see that both our $L$ and
the average value of $\langle |b|\rangle$ obtained from $P(E_{T}, b)$
are in excellent accord with NA50 results in \cite{na50}.  To compare
to NA50, we calculate $L$ for $\rho_0\rightarrow 0.132$~fm$^{-3}$ (a
more realistic value is $0.17$~fm$^{-3}$).  More importantly, we see
that saturation is nearly achieved in the symmetric Pb+Pb system even
when the average $b$ is rather large.
\begin{figure}[htb]
\vskip -2.0truein
\epsfysize=5.0in
\centerline{\epsffile{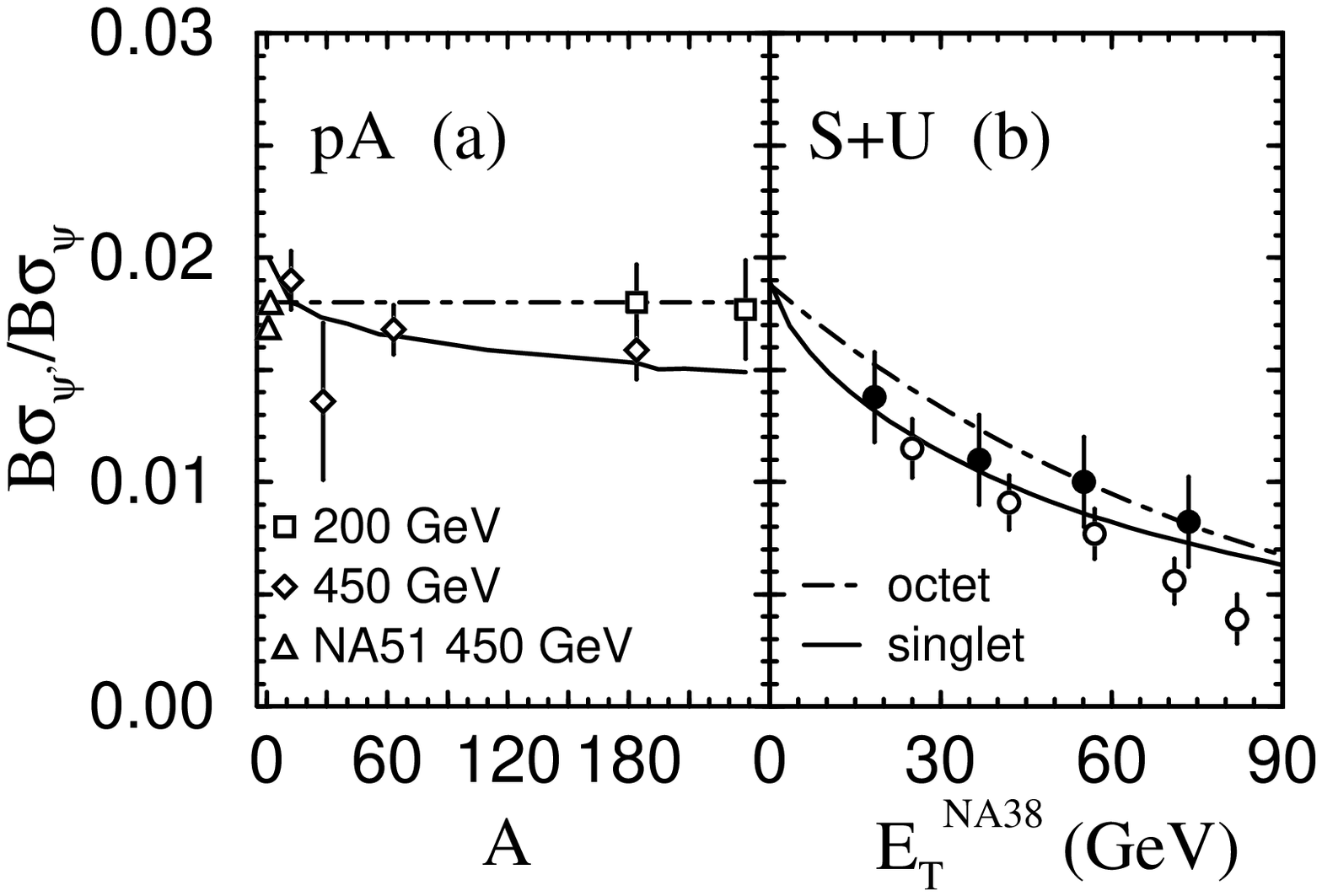}}
\caption{Comover suppression of $\psi^\prime$ compared to (a) NA38 and 
NA51 $p$A data [1, 9] and (b) NA38 S+U data [8] (filled points)
and preliminary data [1].}
\end{figure}

\section{$\psi^\prime$ SUPPRESSION}
\begin{figure}[htb]
\vskip -1.5truein
\epsfysize=6.0in
\centerline{\epsffile{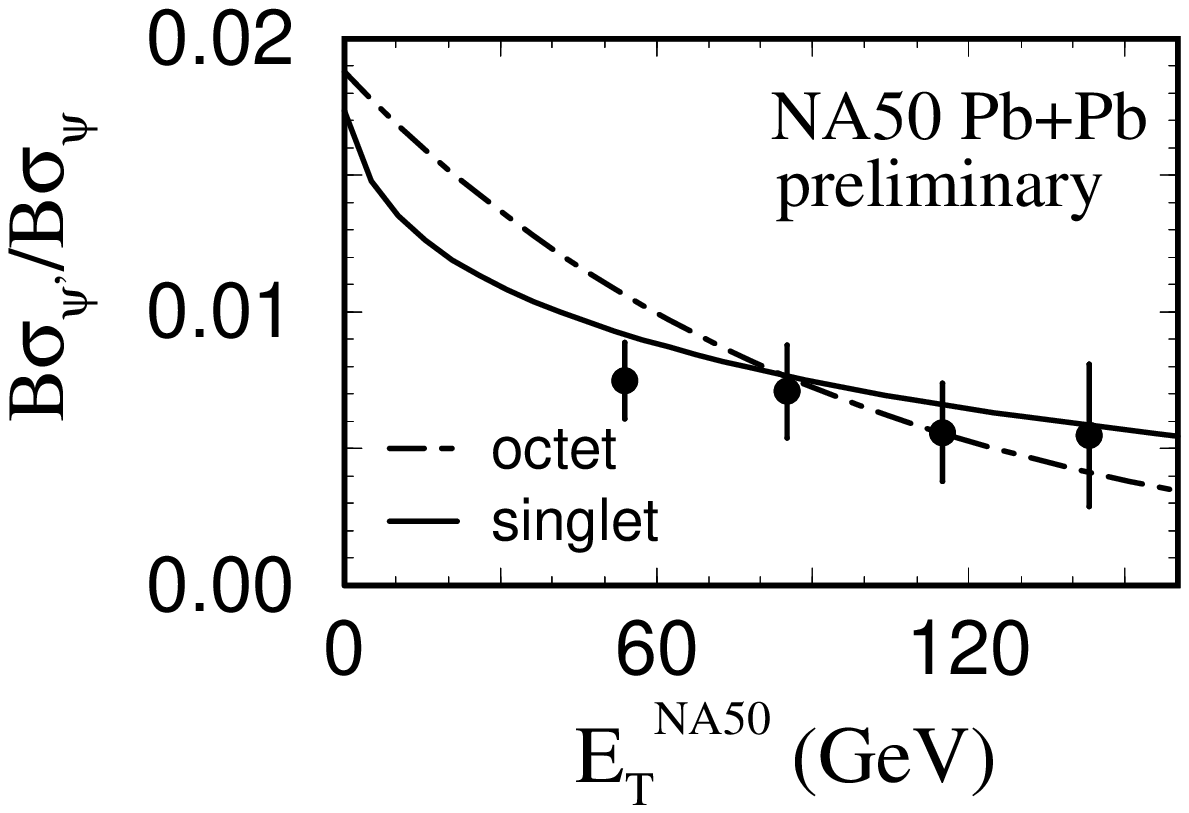}}
\vskip -2.0truein
\caption{Comover suppression in Pb+Pb~$\rightarrow \psi^\prime +X$.}
\end{figure}
To apply eqs.\ (4-6) to calculate the $\psi^{\prime}$--to--$\psi$
ratio as a function of $E_{T}$, we must specify
$\sigma_{\psi^{\prime}}^{NN}$, $\sigma_{\psi^{\prime} N}$, and
$\sigma_{\psi^{\prime} {\rm co}}$.  Following Ref.~\cite{hpc-psi}, we
use $pp$ data to fix $B\sigma_{\psi^{\prime}}^{NN}/B\sigma_{\psi}^{NN}
= 0.02$ (this determines $F_{\psi^\prime}$).  The value of
$\sigma_{\psi^{\prime} N}$ depends on whether the nascent
$\psi^{\prime}$ is a color singlet
hadron or color octet $c\overline{c}$ as it traverses the nucleus. In
the singlet case, one expects the absorption cross sections to scale
with the square of the charmonium radius.  Taking this ansatz and
assuming that the $\psi^\prime$ forms directly while radiative $\chi$
decays account for 40\% of $\psi$ production, one expects
$\sigma_{\psi'}\sim 2.1\sigma_{\psi}$ for interactions with either
nucleons or comovers \cite{gstv}.  For the octet case, we take
$\sigma_{\psi^{\prime} N} \approx \sigma_{\psi N}$ and fix
$\sigma_{\psi^{\prime} {\rm co}}\approx 12$~mb to fit the S+U data.
In fig.~5a, we show that the singlet and octet extrapolations describe
$p$A data equally well.

Our predictions for Pb+Pb collisions are shown in fig.~6.  In the
octet model, the entire suppression of the $\psi^{\prime}$--to--$\psi$
ratio is due to comover interactions.  In view of the schematic nature
of our approximation to $S_{\rm co}$ in eq.\ (8), we regard the
agreement with data of singlet and octet extrapolations as equivalent.

\section{SUMMARY}
In summary, the Pb data \cite{na50} cannot be described by nucleon
absorption alone.  This is seen in the NA50 plot, fig.~1, and
confirmed by our results.  The saturation with $L$ but not $E_T$
suggests an additional density--dependent suppression mechanism.
Earlier studies pointed out that additional suppression was already
needed to describe the S+U results \cite{gstv}; recent data
\cite{na50} support that conclusion (see, however, \cite{bo}).  To
study the S+U question further, experiments can search for saturation
in super--central S+U reactions.  More generally, while comover
scattering explains the additional suppression, it is unlikely that
that explanation is unique.  SPS inverse--kinematics experiments ($B <
A$) and AGS $p$A studies near the $\psi$ threshold \cite{seth} can
help pin down model uncertainties.

In the context of our hadronic scenario, the success of the $\psi$ and
$\psi^\prime$ predictions imply comover densities perhaps exceeding
$\sim 1$~fm$^{-3}$ in Pb+Pb collisions.  Similar densities are found
in string and cascade models of heavy ion collisions.  We must ask if
such hadronic models are applicable at such densities --- they must
certainly be at the very edge of their applicability. Nevertheless,
such models have proven very useful as predictive phenomenological
tools.  The ability to predict hadronic backgrounds that we have
demonstrated is a prerequisite for using charmonium to search for new
phenomena at RHIC and LHC.

We thank Claudie Gerschel and Michel Gonin for discussions of the NA38
and NA50 data and Miklos Gyulassy, Geppetto McLerran and Ezra Pisarski
for insightful comments.


\begin{thebibliography}{9}
%
\bibitem{na50} M.~Gonin {\it et al.} (NA50), Proc. Quark Matter '96,
Heidelberg, Germany, P.~Braun-Munzinger {\it et al.}, eds. (1996).
%
\bibitem{bo} J.-P.\ Blaizot and J.-Y.\ Ollitrault, (1996) hep-ph/9606289;  
D.\ Kharzeev and H.\ Satz, Proc.\ Quark Matter '96, {\it op cit.};
A.\ Capella, private communication. 
%
\bibitem{gv2} S.~Gavin and R.~Vogt, LBL-37980 (1996), hep-ph/9606560.
%
\bibitem{gv} S.~Gavin and R.~Vogt, Nucl.\ Phys.\ B345 (1990) 104.
%
\bibitem{gstv} S.~Gavin, H.~Satz, R.~L.~Thews, and R.~Vogt, Z.\ Phys.\ C61
(1994) 351; S.~Gavin, Nucl.\ Phys.\ A566 (1994) 287c.
%
\bibitem{na38} O.~Drapier {\it et al.} (NA38) Nucl.\ Phys.\
A544 (1992) 209c.
%
\bibitem{na38c}
C.\ Baglin {\it et al.} (NA38) Phys.\ Lett.\ B270 (1991) 105.
%
\bibitem{na38d} C.\ Baglin {\it et al.} (NA38) 
Phys.\ Lett.\ B345 (1995) 617;
S.\ Ramos {\it et al.} Nucl.\ Phys.\ A590 (1995) 117c.
%
\bibitem{na38e} C.\ Lourenco (NA38/NA50), Europhysics Conf.\
on High Energy Physics - EPS-HEP, Brussels (1995).
%
\bibitem{gh} C.\ Gerschel and J.\ H{\"u}fner, Z. Phys. C56 (1992) 171.
%
\bibitem{hpc-psi} R.\ Gavai {\it et al.}, Int.\ J.\ Mod.\ Phys.\ A10
(1995) 3043.
%
\bibitem{hpc-dy} S.\ Gavin {\it et al.}, Int.\ J.\ Mod.\ Phys.\ A10
(1995) 2961.
%
\bibitem{na49} S.\ Margetis {\it et al.} (NA49), 
Phys.\ Rev.\ Lett.\ 75 (1995) 3814.
%
\bibitem{na38f}
M.C. Abreu {\it et al.}, Nucl.\ Phys.\ A566 (1994) 77c.
%
\bibitem{seth}
K.\ Seth, Proc.\ Int.\ Conf.\ on Flavor and Spin in Hadronic Physics,
Torino, eds. F. Balestra {\it et al.} (Editrice Compositori, Bologna,
1994).
%
\end{thebibliography}
\end{document}